\documentclass{sig-alternate}
\usepackage{epsfig,endnotes}
\usepackage{url}
\usepackage{graphicx}
\usepackage{epstopdf}
\usepackage[english]{babel}
\usepackage{float}
\usepackage{epstopdf}
\usepackage{listings}
\usepackage{hyperref}
\usepackage{color}
\usepackage{caption}
\usepackage{setspace}
\usepackage{balance}
\usepackage{amsmath}

\definecolor{brown}{rgb}{0.64,0.16,0.16}
\definecolor{gray}{rgb}{0.5,0.5,0.5}
\definecolor{mauve}{rgb}{0.58,0,0.82}
\newcommand{\mysection}[1]{\vspace{-8pt}\section{#1}\vspace{-4pt}}
\newcommand{\mysubsection}[1]{\vspace{-6pt}\subsection{#1}\vspace{-4pt}}

\makeatletter
\def\@copyrightspace{\relax}
\makeatother
\begin{document}

\title{Adapting Graph Application Performance Via \\ Alternate Data Structure Representations}

\numberofauthors{3}
\author{
\alignauthor
Amlan Kusum\\
       \affaddr{University Of California, Riverside}\\
       \email{akusu001@cs.ucr.edu}       
\alignauthor
Iulian Neamtiu\\
       \affaddr{University Of California, Riverside}\\
       \email{neamtiu@cs.ucr.edu}
\alignauthor
Rajiv Gupta\\
       \affaddr{University Of California, Riverside}\\
       \email{gupta@cs.ucr.edu}
}
\maketitle
\vspace*{-0.5in}

\begin{abstract}
Graph processing is used extensively in areas from social networking
mining to web indexing. We demonstrate that the performance and
dependability of such applications critically hinges on the graph data
structure used, because a fixed, compile-time choice of data structure
can lead to poor performance or applications unable to complete. To
address this problem, we introduce an approach that helps programmers
transform regular, off-the-shelf graph applications into adaptive,
more dependable applications where adaptations are performed via
runtime selection from alternate data structure representations.
Using our approach, applications dynamically adapt to the input
graph's characteristics and changes in available memory so they
continue to run when faced with adverse conditions such as low memory.
Experiments with graph algorithms on real-world (e.g., Wikipedia
metadata, Gnutella topology) and synthetic graph datasets show that
our adaptive applications run to completion with lower execution time
and/or memory utilization in comparison to their non-adaptive
versions.
\end{abstract}

\vspace{-0.1in}
\keywords
\vspace{-0.05in}
runtime data structure selection, space-time trade-off


\mysection{Introduction}
Graph processing continues to increase in popularity with the
emergence of applications such as social network mining, real-time
network traffic monitoring, etc. Due to their
data-intensive nature, the performance and dependability of such applications
depends upon how well the choice of runtime data structure matches the
input data characteristics and availability of memory (low memory can prevent the applications from completing). 


\textsf{Input Data Characteristics.}
Programmers often choose specific, fixed data structures when developing graph applications. The memory used by the data structure can be greatly influenced by the input data characteristics. Thus, it is possible that the characteristics of data may not match the choice of the data structure. This is particularly problematic when the application is expected to encounter a wide range of input data characteristics, and these characteristics may change during the course of execution. For example, matrices can be represented in the Compressed Column Storage (CCS) format, appropriate for sparse matrices, or the array representation, appropriate for dense matrices. An application, e.g., matrix multiplication, programmed to use the sparse CCS format, could take longer to complete when presented with a dense input. Similarly, evolving graphs~\cite{konnect}, where nodes or edges are added during execution, are another example of changes in input data characteristics. The data structure selection based on input pre-analysis will fail under such scenario.  Therefore, in our approach, \emph{adaptive applications tailor the choice of data structure to match input data characteristics at runtime.}

\textsf{Availability of Memory.}
Since real-world applications often do not run in isolation, they share the available memory resources with other applications. There could be times where the application experiences a resource crunch, caused by other running programs. In this scenario the performance of the application may be degraded, or the application may even be prematurely terminated. Therefore, in our approach,
\emph{adaptive applications tailor the choice of data structure to match availability of memory at runtime.} 

It is well known that for data-intensive applications, the choice of data structure is critical to memory usage and execution time. There has been previous work on data structure identification~\cite{ddt}, as well as data structure prediction and selection~\cite{brainy,sharir,r17,r27}. While these prior approaches help in data structure selection, none of them support switching from one data structure to another as the application executes. There has also been work on dynamically adapting the representation of individual data items for impacting memory usage and performance---employing data compression~\cite{compress} or replacing {\tt float} data with {\tt int} data~\cite{elastin}. These techniques are orthogonal to our work that switches between alternate high level data structures. Other approaches dynamically switch between implementations. Elastin~\cite{elastin} allows a program to switch between versions using dynamic software update techniques \cite{ginseng1, ginseng2}; however, it does not consider switching between alternate high level data structures. IBM's K42 Operating System~\cite{k421,k423} supports hot-swapping classes as a mechanism for performing dynamic updates. Scenario Based Optimization~\cite{sbo}, a binary level online optimization technique dynamically changes the course of execution through a route meant for a particular runtime scenario as predefined by developer. Wang et al.~\cite{react} proposed dynamic resource management techniques based on user-specific, application-specific and hardware-specific management policies. In contrast, our objective is to simultaneously support alternate data structures and switch between them.

In this paper we consider several widely-used graph applications and study how data structure representations impact execution time and memory consumption on a range of input graphs (Section~\ref{sec_motivation}). 
The input graphs consist of both real-world graphs such as Wikipedia metadata, Gnutella network topology (from the SNAP library~\cite{snap}), and synthetic graphs. Based upon the observations from our study, we design a concrete adaptation system that supports switching between alternate representations of the data in memory (Section~\ref{sec_approach}). We demonstrate that the cost of performing the runtime adaptations is quite small in comparison to the benefits of adaptation (Section~\ref{sec_Eval}). Moreover, the lightweight monitoring we employ to detect adaptation opportunities imposes acceptable overhead even when no adaptations are triggered at runtime. Thus, our adaptive versions have nearly the same performance as the most appropriate non-adaptive versions for various input characteristics. We compare our approach with related work in Section~\ref{sec_Rel}, and in Section~\ref{sec_Conc} we conclude.
\vspace{0.03in}
\vspace{-0.2in}
\section{A Study of Graph Applications}
\label{sec_motivation}
In this section we study the execution time and memory usage behavior of a set of graph applications. The goal of this study is two fold. First, we want to quantify how input data characteristics and the choice of data structures used to represent the graphs impact memory usage and execution time. Second, we would like to develop a simple characterization of program behavior that can be used to guide data structure selection at runtime.

We considered six graph algorithms: Muliple Source Shortest Path (MSSP) finds the shortest path from all the nodes to every other node; Betweenness Centrality (BC) computes the importance of a node in a network;
Breadth First Search (BFS) traverses the graph with each node as root per iteration; Boruvka's Algorithm (MST-B) and Kruskal's Algorithm (MST-K), finds the minimum spanning tree; Preflow Push (PP), finds out the maximum flow in a network starting with each individual node as source. The core data structure used in these applications is a graph. We consider two different representations of graphs: Adjacency List (\texttt{ADJLIST}); and Adjacency Matrix (\texttt{ADJMAT}). When the graph is sparse, it is expected that \texttt{ADJLIST} will use less memory than \texttt{ADJMAT}. On the other hand, for highly dense graphs \texttt{ADJMAT} may use less memory than \texttt{ADJLIST}. Determining whether a pair of nodes is connected by an edge can be done in constant time using \texttt{ADJMAT} while it may require searching through a list with \texttt{ADJLIST}. Thus, the runtime memory usage and execution time depend upon the sparsity, or conversely the density, of the input graph. The input graphs with relevant properties and densities were generated to study program behavior.

To observe the trade-offs of using the alternative representations of graphs, we executed each of the programs using the two representations. The programs were run on inputs consisting of randomly-generated graphs with varying density which is computed as $\frac{\left|E\right|}{\left|V\right| \left|\left(V-1\right)\right|}$, where $\left|V\right|$ and $\left|E\right|$ are number of nodes and edges in the graph. The inputs were selected such that the trade-offs could be exposed easily. The results of these executions are summarized as follows:

%
\vspace{0.1in}
\begin{table}
\centering
{\small
\caption{Relative performance ranges.}
\begin{tabular}{|l|c|c|}
\hline 
Application & \multicolumn{2}{c|}{ADJLIST / ADJMAT}  \\ \cline{2-3}
& Memory Usage & Execution Time  \\
\hline \hline
MSSP &  0.68 - 8.02 & 0.40 - 4.00    \\ 
\hline 
BC   &  0.40 - 4.00 & 0.59 - 2.88 \\ 
\hline 
MST-K &  0.72 - 2.44 & 0.35 - 3.83 \\ 
\hline 
BFS & 0.54 - 5.47 &  0.72 - 3.14 \\ 
\hline 
MST-B & 0.71 - 1.67 & 0.16 - 7.21 \\ 
\hline 
PP & 0.60 - 5.40  & 0.50 - 3.53 \\ 
\hline 
\end{tabular} 
\label{tbl:perfranges}
}

\vspace{0.1in}

\centering
{\small
\caption{Density ranges where each data structure prevails.}
\begin{tabular}{|p{1.7cm}|p{1.7cm}|p{1.8cm}|p{1.7cm}|}
\hline 
 
 Application & ADJLIST best space \& time & ADJLIST best space, ADJMAT best time &  ADJMAT best space \& time \\ 
\hline 
MSSP & $<$ 9\%  &  9\% - 25\% & $>$ 25\% \\ 
\hline 
BC & $<$ 10\% & 10\% - 25\%  & $>$ 25\%\\ 
\hline 
MST-K & $<$ 25\%  & 25\% - 37\% & $>$ 37\%\\ 
\hline 
BFS & $<$ 8\% & 8\% - 25\% & $>$ 25\%\\ 
\hline 
MST-B & $<$ 10\% & 10\% - 40\% & $>$ 40\% \\ 
\hline 
PP & $<$ 2\% & 2\% - 34\% & $>$ 34\%\\ 
\hline 
\end{tabular} 
\label{dsranges}		
}

\vspace{-0.2in}
\end{table}

\vspace{-0.075in}
\textsf{Impact of data structure selection on memory usage and execution time.} We present the relative memory usage and execution time of program versions in Table~\ref{tbl:perfranges}. In particular, we computed the ratios of memory usages and execution times for \texttt{ADJLIST} and \texttt{ADJMAT} versions across all graph densities considered. The minimum and maximum values of observed ratios is given in Table~\ref{tbl:perfranges}. As we can see, in terms of both memory usage and execution time, the relative performances vary a great deal. Moreover, neither representation gives the best memory usage or execution time performance across all graph densities. Hence, it is crucial to select the data structure at runtime, based upon the input data characteristics. 

\textsf{Characterization of application behavior.} For the purpose of runtime data structure selection, we characterize the behavior of each application as shown in Table~\ref{dsranges}. Note that graph densities are divided into three subranges. In the first range (e.g., $< 9\%$ for MSSP) the \texttt{ADJLIST} is both more memory- and time-efficient than \texttt{ADJMAT}. In the second range (e.g., $9\% - 25\%$) \texttt{ADJLIST} is more memory-efficient while \texttt{ADJMAT} is more time-efficient. Thus, the selection can be made at runtime based upon memory availability. Finally, in the third range (e.g., $>25\%$ for MSSP) \texttt{ADJMAT} is both more memory and time efficient than \texttt{ADJLIST}.

\vspace{-0.025in}
\mysection{Adaptive Applications}
\label{sec_approach}

\begin{figure}[h]
\centering

\includegraphics[width=.375\columnwidth, angle=270]{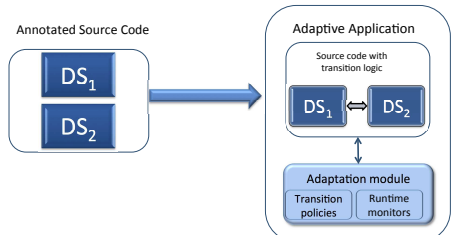}
\vspace{-0.01in}
\caption{High level overview of our approach.}
\vspace{-0.175in}
\label{fig:framework}
\end{figure}

We now present our approach for building adaptive applications; an overview is shown in Figure~\ref{fig:framework}. The starting point is the \emph{annotated source code}: in the source code, programmers add \emph{annotations} to identify the alternative data structures, e.g., DS$_1$ and DS$_2$, and functions operating on them. The compiler takes heed of these annotations and generates the \emph{source code with transition logic}, that is capable of dynamically switching among alternative data structure representations. The transitions are allowed at selected program points where the processing of an input item has just completed and that of another item is about to begin. Lastly, the \emph{adaptation module} consists of the runtime monitors for tracking input data characteristics and memory usage as well as the code that implements the transition policy that triggers the switch from one data structure representation to another. The adaptation can be triggered by a mismatch between the input data characteristics and the data structure currently in use. To discover this mismatch the characterization of application behavior as performed in the previous section is used. The adaptation can also be triggered by the system during high memory usage. 


\textsf{Programming for adaptation.}
To enable adaptation, the programmer implements the alternate data structures. In addition, a compute-intensive function during whose execution adaptation may be performed, must be coded as follows. First, it should contain a variable that tracks the progress in terms of processing steps defined as either the amount of input processed or results produced. Second, it should be written so that it can commence execution from any point between two processing steps. The latter is needed because we allow execution to switch from one data representation to another at these points. We used a set of pragmas in our approach to identify alternate data structure representations, enable generation of code that transfers code from one representation to another, and identify program points where transitions may be performed. First, the programmer identifies the data structure to the compiler. The programmer annotates the alternate representation of data structures in multiple files with \lstinline|#pragma ADP(<SRC_FILENAME>, "data1_def")|. \lstinline|<SRC_FILENAME>|'s presence clearly differentiates the alternate representation of the data structure in multiple files. If there are multiple data structures with alternate representations in different files, then they could be annotated with a different index, e.g., \lstinline|#pragma ADP(<SRC_FILENAME>, "data2_def")|. Second, the programmer uses several pragmas to identify the key methods (insert, delete, traverse, and fetch) that manage data stored in the data structure. Another pragma allows access to the initialization parameters which must be migrated from one data structure to another. All of this information is used to generate the code for data and function migration when we switch between data structures.

%
%

\textsf{Triggering adaptations.} 
The adaptation module decides whether or not to switch between data structures based upon the input from runtime monitors and the transition policy. Since the adaptation could be program-triggered or system-triggered, there are two kinds of monitors which are required by the adaptation module. The input data monitor captures input data characteristics and the memory monitor reports the available system memory. The transition policy defines which data structure representation is better for what range of input data characteristics in terms of execution time and memory consumption. Its specification consist of three parts, as illustrated below:
\vspace{-0.05in}
\begin{lstlisting}
       /* EXECUTION TIME */
           DS1 [0,9) 
           DS2 [9,100]
      /*MEMORY*/
           DS1 [0,25) 
           DS2 [25,100]
     /*THRESHOLD*/
           MEMORY 100
\end{lstlisting}
\vspace{-0.05in}

The first part indicates the ranges for which a particular data structure representation is best in terms of execution time: under \lstinline|EXECUTION TIME| in the figure, the input data property for which \texttt{ADJLIST} (DS1) is better is denoted by directives \lstinline|DS1|, which means that \texttt{ADJLIST} is favorable in terms of execution time if the input data property or density of the graph (in case of MSSP) is in between 0\% and 9\%. The second part consists of the ranges of the input data property for which a particular data structure representation is better in terms of memory. According to the figure, under \lstinline|MEMORY|, we see that \texttt{ADJLIST} (DS1) is better when the density of the input graph is between 0\% and 25\% while \texttt{ADJMATRIX} (DS2) is better when the density of the graph is between 26\% and 100\%. The third part is the threshold for memory, defined by the programmer to notify the system that if the available memory is below this threshold then, regardless of input data characteristics always use the representation requiring least memory; in the figure (under \lstinline|THRESHOLD|) the threshold is set to 100MB. 

\begin{figure}[ht]
\vspace{-0.2in}
\centering
\begin{lstlisting}
dataMigrationDS1DS2(void* DS1, void* DS2)
{
  initializationParameters* ip;
  ip = getInitializationParameter(DS1);
  initializeDS2(&DS2,ip);
  transferDataDS1DS2(&DS1,&DS2)
  deleteDS1(&DS1);
}
transferDataDS1DS2(void** DS1, void** DS2)
{
  i = 0; void* dataValue;
 for(i = 0;i< **DS1->maxData;i++) {
    dataValue = fetchDataDS1(i,*DS1);
    if(dataValue != NULL) {
      insertDataDS2(*DS2, dataValue, i);deleteDataDS1(i,*DS1);
}}}
\end{lstlisting}
\vspace{-0.2in}
\caption{Data migration.}
\vspace{-0.2in}
\label{fig:src_datamigration}
\end{figure}

\begin{figure}[t]
\centering
\begin{tabular}{|p{3.8cm}|p{3.8cm}|}\hline
\vspace{-0.15in}
\begin{lstlisting}
#pragma ADP("DS1",
		"ds1_op1") 
void computeMSSP_DS1(
         void* graph, void* rs,
         int* progress);
...
  computeMSSP_DS1(graph, 
          rs, progress);
 
...
\end{lstlisting}
\vspace{-0.2in} 
 &
 \vspace{-0.15in}
\begin{lstlisting}
//#pragma ADP("DS1",
		"ds1_op1")
void computeMSSP_DS1(
          void* graph,void* rs, 
          int* progress);
...
  callOP1(graph,
          rs,progress, 
          startDS);
... 
\end{lstlisting} \vspace{-0.2in}  \\ \hline

\end{tabular}
\vspace{0.05in}
\caption{Function migration method before compilation (left) and after compilation (right).}
\label{fig:code1}
\vspace{-0.25in}
\end{figure}

\textsf{Switching between data structure representations.} 
The data structure transition logic is inserted into the source files by the compiler, guided by the pragmas. This transition logic carries out on-the-fly transitions from one data structure representation to another whenever required. To accomplish the transition, the in-memory data must be transformed from one representation to another, along with the functions operating on them. The transition logic handles this by function migration and in-memory data methods contained in the logic. When the code for transition logic is inserted, appropriate header files are also inserted such that source code after modification compiles and links properly. To avoid recomputation of already-computed results, the result transfer logic (injected into code along with the transition logic) will transfer the already-computed results from one representation to the other representation.

\lstset{numbers=left}
\begin{figure}[t]
\begin{lstlisting}[xleftmargin=.2in]
void callOP1(void* ds, void* rs, int progress, currentDS){
  extern int changeReq; void* newDS; void* newRS;
  while(progress < 100){
    if(changeReq == 1){ switch(currentDS) {
        case 1:         
          currentDS = 2; dataMigrationDS1DS2(ds, newDS);
          resultMigrationRS1RS2(rs, newRS);
          ds = newDS; newDS = NULL;  rs = newRS;  newRS = NULL;
          computeMSSPDS2(ds, rs, progress);
          break;
        case 2:
          currentDS = 1; dataMigrationDS2DS1(ds, newDS);
          resultMigrationRS2RS1(rs, newRS);
          ds = newDS; newDS = NULL; rs = newRS; newRS = NULL;
          computeMSSPDS1(ds, rs, progress);
          break;            
      }}
    else { switch(currentDS) {
        case 1: computeMSSPDS1(ds, rs, progress);  break;
        case 2: computeMSSPDS2(ds, rs, progress);  break;   
  }}}}
\end{lstlisting}
\vspace{-0.1in}
\caption{Switching between implementations.}
\label{fig:src_fnmigration}
\vspace{-0.3in}
\end{figure}

An example data migration function is shown in Figure~\ref{fig:src_datamigration}. The code in the figure transfers the data from the data structure representation DS1 to another representation DS2. It begins with initialization of the DS2 data structure representation. The initialization parameters are fetched from DS1 and they consist of standard parameters that are invariant in both DS1 and DS2. For example, in the MSSP benchmark the invariant data is the number of nodes. In the PP benchmark the invariant data consists of number of nodes, the height, capacity and flow of each node. The \lstinline|transferData| function is generated from \lstinline|traverseData| function of DS1 as provided by the developer. This function traverses through the data by reading each data value, migrating it to DS2 representation using \lstinline|insertDataDS2| and also deleting that data from DS1 using \lstinline|deleteDataDS1| thus releasing memory. The \lstinline|deleteDS1| clears memory which contains the data regarding the initialization parameters. 

The transition between implementations, i.e., switching from one set of functions operating on representation DS1 to functions operating on representation DS2 must be carefully orchestrated. The developer denotes an operation with a directive such as \lstinline|#pragma ADP("DS1","data1_op1")|, which informs the compiler that the function is compute-intensive, as shown in Figure~\ref{fig:code1}. Any call to that function is replaced by our customized method, which checks and executes operations with the suitable data structure. In this example \lstinline|computeMSSP_DS1| is replaced by \lstinline|callOP1|. The additional parameter, \lstinline|startDS|, denotes the type of the current data structure representation in memory. The other three parameters are the data structure, a progress gauge, and the result set for storing the result. For example in the case of MSSP, a method that finds MSSP has the signature \lstinline|void computeMSSP_DS1(void* graph, void* rs ,int* progress)|. The first parameter is the input graph and the second parameter \lstinline|rs| stands for the result set and its declaration must be annotated by the programmer with \lstinline|#pragma ADP("DS1", "data1_res1")|. The last parameter identifies the progress, which is the iteration number of the outer most long running loop. For example, if the method is called with a progress value 10, then the execution is started from progress value 10 and continuously updated with the loop iteration number.

\lstset{numbers=none}
\begin{figure}[t]

\begin{tabular}{|p{3.8cm}|p{3.8cm}|}\hline 
\vspace{-0.15in}
\begin{lstlisting}
void computeMSSP_DS1(
         void* graph,
         void* rs, 
         int* progress){
...
#pragma ADP("DS1",
	"ds1_op1_safe")
...



}


\end{lstlisting}
\vspace{-0.2in}
 &
 \vspace{-0.15in}
\begin{lstlisting}
void computeMSSP_DS1(
          void* graph,
          void* rs, 
          int* progress){
...
//#pragma ADP("DS1"
	,"ds1_op1_safe") 
  if(checkChangeStatus()==1) {
    *progress = curProgress; 
    return;
  }
}
\end{lstlisting} \vspace{-0.25in} \\  \hline
\end{tabular}
\vspace{0.025in}
\caption{Adaptation module interrupt before compilation (left) and after compilation (right).}
\label{fig:code2}
\vspace{-0.18in}

\end{figure}

The detailed function selection and migration activity is shown in Figure~\ref{fig:src_fnmigration}---for MSSP benchmark. An external variable \lstinline|changeReq|, set by the adaptation module, is checked (line~4). If a transition has been requested, then first the data is migrated from one data structure representation to another (lines~6 and~12). Next, if needed, the result is migrated from one representation to another (lines~7 and~13). Finally, the corresponding MSSP function for that data structure is called (lines~9 and~15) and the operation is resumed from the progress point. If there is a change request from the adaptation module, then operation is paused and it returns back to \lstinline|callOP1|. This process continues until the MSSP computation completes. 

The question arises where ongoing MSSP computations should be interrupted to check if the adaptation module has requested a change or not. To solve this problem, we rely on the programmers to use the directive \lstinline|#pragma ADP("DS1", "ds1_op1_safe")| to indicate the safe transition points in \lstinline|operation1| as shown in Figure~\ref{fig:code2}. This directive notifies our framework that, if the operation is paused and the transformation is performed at that point, then there is minimal recomputation of result. This is typically the end of an iteration in long-running loops. Since the programmer is well aware of the long running loops in the compute-intensive function, it is best to have the programmer mark the points appropriate for the insertion of adaptation module interrupts. The directive is replaced by an interrupt which checks if there is a change required and thus returns back to \lstinline|callOP1|.

\begin{table}
\vspace{4pt}
\centering
\caption{Comparison of execution times of non-adaptive versions with adaptive version under program triggered adaptations, on the original {\tt p2p-Gnutella} graph. Note that \texttt{ADJLIST} is the better representation.}
\hspace{-0.15in}
{\scriptsize
\begin{tabular}{|l|p{1cm}|p{1.1cm}|p{1.1cm}|p{1.2cm}|p{1.1cm}|}
\hline 
\!\!\!App.\!\!\!  & \multicolumn{2}{p{2.2cm}|}{Non-Adaptive Ex. Time (sec)} & \multicolumn{2}{p{2.2cm}|}{Adaptive: ADJMAT$\rightarrow$ ADJLIST (sec) \!\!\!} & Benefit Realized \\ \cline{2-5}
 & \!\!\!ADJLIST  & \!\!\!ADJMAT & Ex. Time & Transition Latency & \!\!\!(\%)\!\!\!  \\ \hline \hline
MSSP  &  1,386 & 2,489  & 1,408 &  3.00   & 98.05 \\  \hline 
BC    & 1,362 & 2,565 & 1,383  &  2.73   & 98.32 \\ \hline 
MST-K  &  389 &  956 &   397   &  3.17   & 98.55 \\ \hline 
BFS   &  1,434 & 2,594  & 1,457   &  2.56  & 98.05 \\\hline 
MST-B  &  1,454 & 2,114 & 1,465   &  3.15    & 98.35 \\ \hline 
PP  & 81  &  256  & 87  &  3.2  & 96.35  \\ 
\hline 
\end{tabular}
} 

\vspace{-0.25in}
\label{pt-table}
\end{table}
\begin{figure}[h]
\vspace{-0.1in}
\centering
\hspace{-0.05in}
\includegraphics[width=9cm]{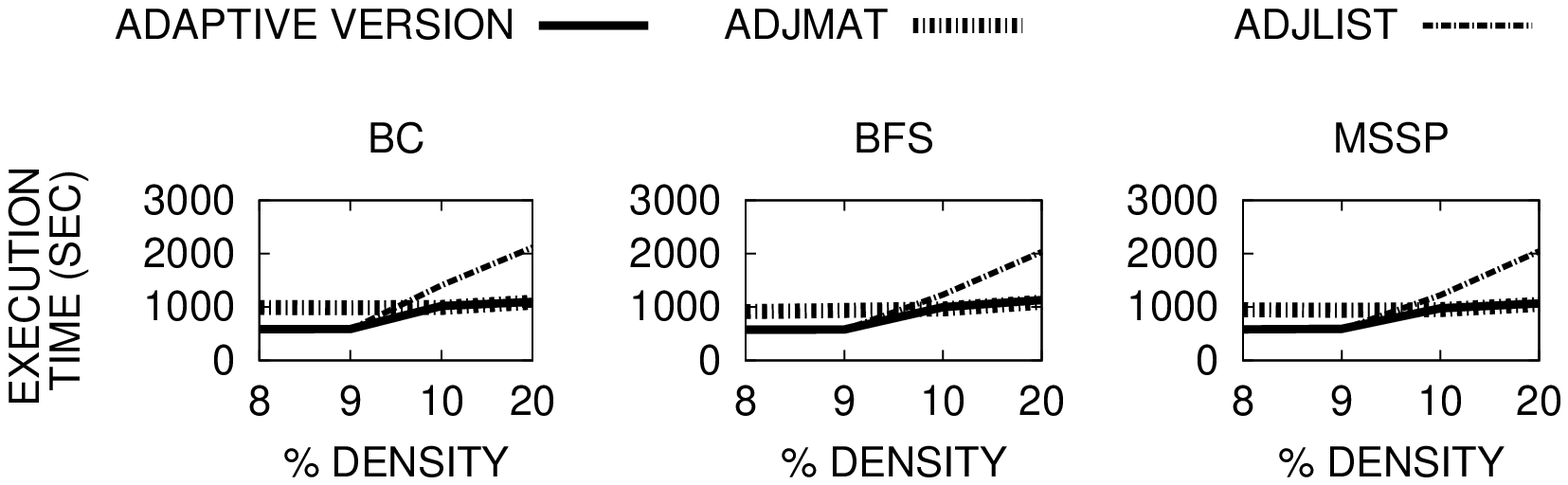}
\hspace{-0.05in}
\includegraphics[width=9cm]{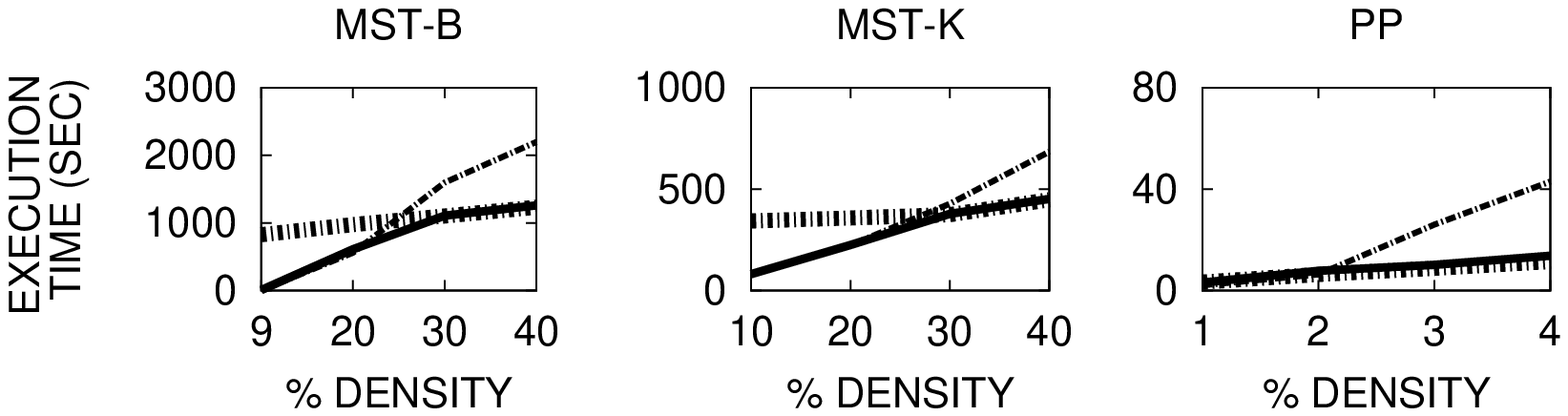}
\caption{Adaptive vs. non-adaptive performance.}
\label{overall}
\vspace{-0.1in}
\end{figure}

\vspace{-0.1in}
\mysection{Evaluation}
\label{sec_Eval}
In this section we evaluate the performance of adaptive versions of graph algorithms and compare them with corresponding non-adaptive versions of the applications. The goals of these experiments are as follows. First, we evaluate the efficiency of our approach by measuring its benefits and overhead. Second, we consider the benefits of adaptation under two scenarios: adaptation triggered by the input characteristic, i.e., graph density; and system triggered adaptation.  All experiments were run on a 24-core machine (4 six-core AMD Opteron$^{TM}$ 8431 processors) with 32GB RAM. The system ran Ubuntu 10.04, Linux kernel version 2.6.32-21-server. The sources were compiled with Gcc 4.4.3. 

\emph{Real World Data-sets:}
We evaluate our system on some of the real-world graphs from the SNAP graph library~\cite{snap}. 
The first graph, {\tt wiki-Vote}, contains the who-votes-for-whom graph
in Wikipedia administrator elections. This graph has 7,115 nodes and
103,689 edges. The second graph, {\tt p2p-Gnutella}, is a snapshot of
Gnutella, a decentralized peer to peer file sharing network from
August 9, 2002. This graph has 8,114 nodes representing hosts and
26,013 edges representing the connections between these hosts. For
experiments, in cases where a more dense graph was needed, we
added edges in both the graphs to raise the required density. 
\vspace{-0.1in}
\mysubsection{Programming Effort} 
\vspace{0.04in}
The programmers need to add annotations to transform off-the-shelf applications to adaptive ones. In addition to this, programmers also need to modify the compute-intensive methods so they can be executed in incrementalized fashion. The number of pragmas added and the number of additional lines of code added to modify the methods are shown in Table~\ref{tbl_preffort}. As we can see, these numbers are fairly modest.


\begin{table}
\centering
\caption{Programming effort.}
\label{tbl_preffort}
{\small
\begin{tabular}{|p{1.6cm}|c|c|c|c|c|c|c|}
\hline
Application & MSSP & BC   & \!\!MST-K\!\!& BFS  & \!\!MST-B\!\!&  PP  \\ \hline
\# pragmas & 8 & 9 & 9  & 8 & 9  & 8 \\ \hline
Additional LOC & 9 & 12 & 10  & 8 & 14  & 6 \\ \hline
\end{tabular}
}
\vspace{-0.22in}
\end{table}


\vspace{-0.025in}
\subsection{Input Triggered Adaptation}
\vspace{-0.025in}
In this scenario we study how adaptive applications respond to the mismatch between the data structure representation fixed a priori at compile time and the density of the input graph. We compute the benefit realized by our approach for various applications. In particular, we start the program by using the \texttt{ADJMAT} representation and select a real world graph {\tt (p2p-Gnutella)} which is 0.004\% dense, which makes \texttt{ADJLIST} the ideal representation. Therefore, when the adaptive application is run, it dynamically switches from the \texttt{ADJMAT} to the \texttt{ADJLIST} representation. 

In Table~\ref{pt-table} we present the \emph{execution times} of the non-adaptive (\texttt{ADJLIST} and \texttt{ADJMAT} representations) and adaptive (\texttt{ADJMAT}$\rightarrow$\texttt{ADJLIST}) versions of the applications. For the latter version, we also present the \emph{transition latency} which is the execution time after which the program has completed the transition to the \texttt{ADJLIST} representation. From the results in Table~\ref{pt-table}, we observe the following. The execution time of the adaptive version, on average, is 2.49\% higher than the non-adaptive \texttt{ADJLIST} version; but 48.09\% lower than the non-adaptive \texttt{ADJMAT} version. For example, for MSSP, the execution of the adaptive version is 1,408 seconds which is 1.54\% higher than the execution time of the non-adaptive \texttt{ADJLIST} version (1386 seconds) and 56.55\% lower than the execution time of the non-adaptive \texttt{ADJMAT} version (2,489 seconds). In addition, we observe that the transition latency of the adaptive version is small in comparison to the total execution time. For example, for MSSP, the transition latency of 3 seconds is approximately 0.21\% of the total execution time of 1,408 seconds. That is, the adaptation is performed quickly (low transition latency) and efficiently (low transition overhead). Thus, nearly all the benefits of using \texttt{ADJLIST} over \texttt{ADJMAT} are realized by the adaptive version. 

We quantify the benefit realized by our approach as follows. The maximum possible benefit is given by the difference in the execution times of the non-adaptive \texttt{ADJMAT} and non-adaptive \texttt{ADJLIST} versions. The benefit our approach realizes is the difference between the execution times of the non-adaptive \texttt{ADJLIST} version and the adaptive version. The realized benefit, as a percentage of maximum possible benefit, is given in the last column of Table~\ref{pt-table}. As we can see, the realized benefit is over 96\% for these applications. 


\begin{table}
\centering
\caption{Breakdown of adaptation overhead.}
\label{tbl_breakdown}
{\small
\hspace{-0.3in}
\begin{tabular}{|p{2.3cm}|c|c|c|c|c|c|c|}
\hline
Application & MSSP & BC   & \!\!MST-K\!\!& BFS  & \!\!MST-B\!\!&  PP  \\ \hline
DS Conversion & 1.56 & 1.44 & 1.98  & 1.4 & 1.98  & 1.98 \\ \hline
Monitoring \& Transition Logic & 10.78 & \!\!10.11\!\! & 5.89  & 12.74 & 7.19  & 2.92 \\ \hline
Suboptimal Mode & 9.07 & 8.62 & 0.3  & 8.41 & 1.71  & 1.48 \\ \hline
\end{tabular}
}
\vspace{-0.2in}
\end{table}

\begin{figure}[!ht]
\centering
\hspace{-0.25in}
\includegraphics[width=9cm]{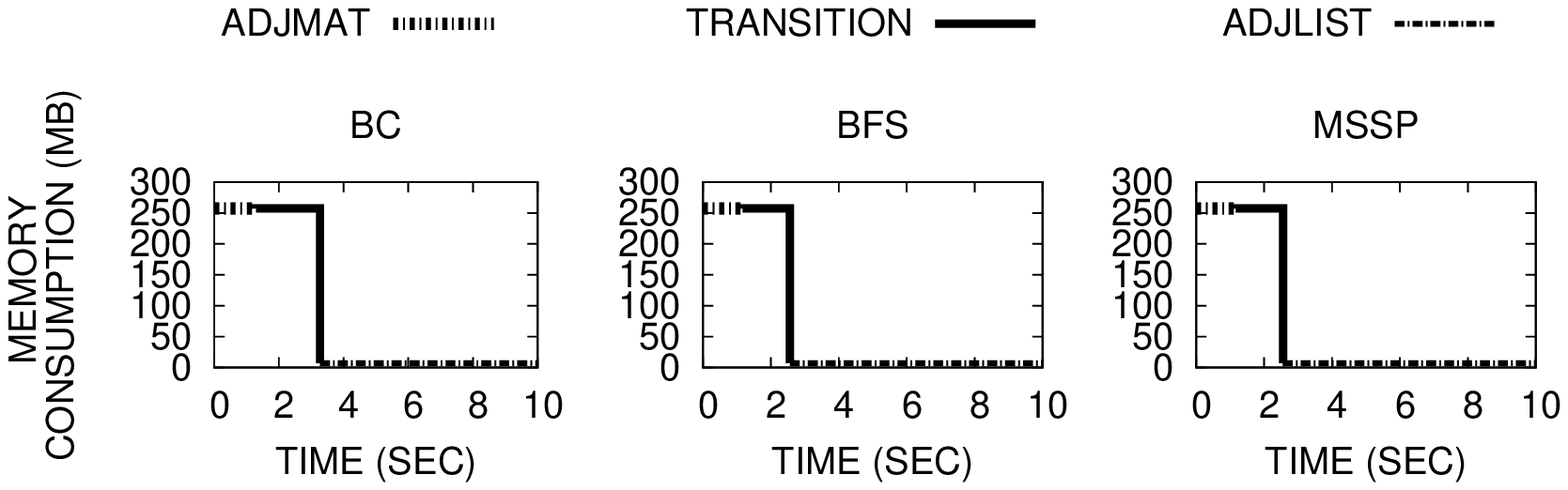}
\hspace{-0.25in}
\includegraphics[width=9cm]{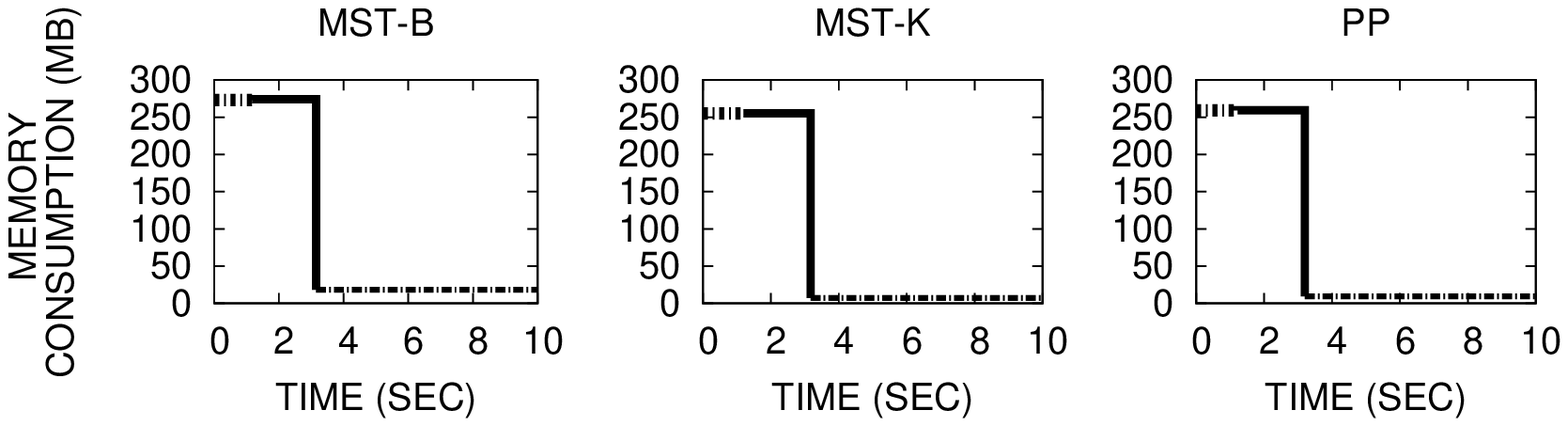}
\caption{Program-triggered adaptation.}
\label{pt-mem}
\vspace{-0.2in}
\end{figure}

The additional execution time taken by the adaptive version over the non-adaptive \texttt{ADJLIST} version can be divided into three categories: time spent on converting from one data structure representation to another; time spent on runtime monitoring and transition logic to trigger adaptation; and the time lost due to running the application in suboptimal mode, i.e., with the \texttt{ADMAT} data structure. The breakdown of the extra execution time into the three categories is shown in Table~\ref{tbl_breakdown}. As we can see, the majority of the time is spent on runtime monitoring and transition logic. The next significant component is the time spent due to running the program in the suboptimal configuration before the transition occurs. Note that the time spent on converting one data structure into another (column 2) is the least. 

An intuitive way to visualize adaptation is to plot how the memory used by applications varies before, during, and after adaptation.
In Figure~\ref{pt-mem} we show how memory ($y$-axis) varies over time ($x$-axis)
when starting the application in the \texttt{ADJMAT} representation and then through adaptation, the application transitions to \texttt{ADJLIST}. The charts point out several aspects.
First, since we are using sparse graphs, as expected, the memory used is reduced significantly (tens of megabytes) when we switch from the \texttt{ADJMAT} to \texttt{ADJLIST} representation. Second, the switch from one data structure to the other takes place fairly early in the execution of the program. Third, the time to perform adaptation and the extra memory used during adaptation are very low.

In Figure~\ref{overall} we show the execution time of the adaptive version for varying input densities over the range where we expect the adaptive application to switch from the \texttt{ADJLIST} to the \texttt{ADJMAT} representation. For these experiments, we have used graph size of 4000 nodes and varied densities. The execution times of the non-adaptive versions that use fixed representations (\texttt{ADJLIST} and \texttt{ADJMAT}) are also shown. As we can see, the performance of the adaptive application is very close to the best of the two non-adaptive versions.

\begin{figure}[!ht]
\vspace{0.1in}
\centering
\includegraphics[width=7cm]{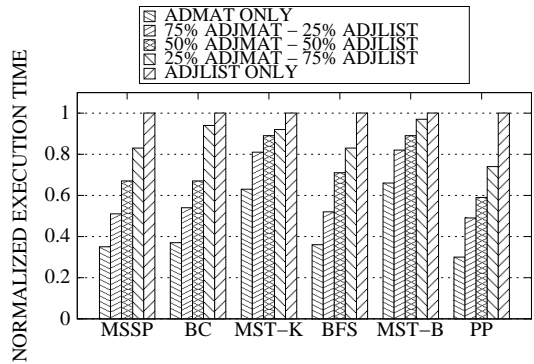}
\vspace{-0.15in}
\caption{Comparison of adaptive vs. non adaptive normalized execution time in system-triggered adaptation.}
\label{st-table}
\vspace{-0.1in}
\end{figure}

\begin{figure}[!ht]
\centering
\hspace{-0.25in}
\includegraphics[width=9cm]{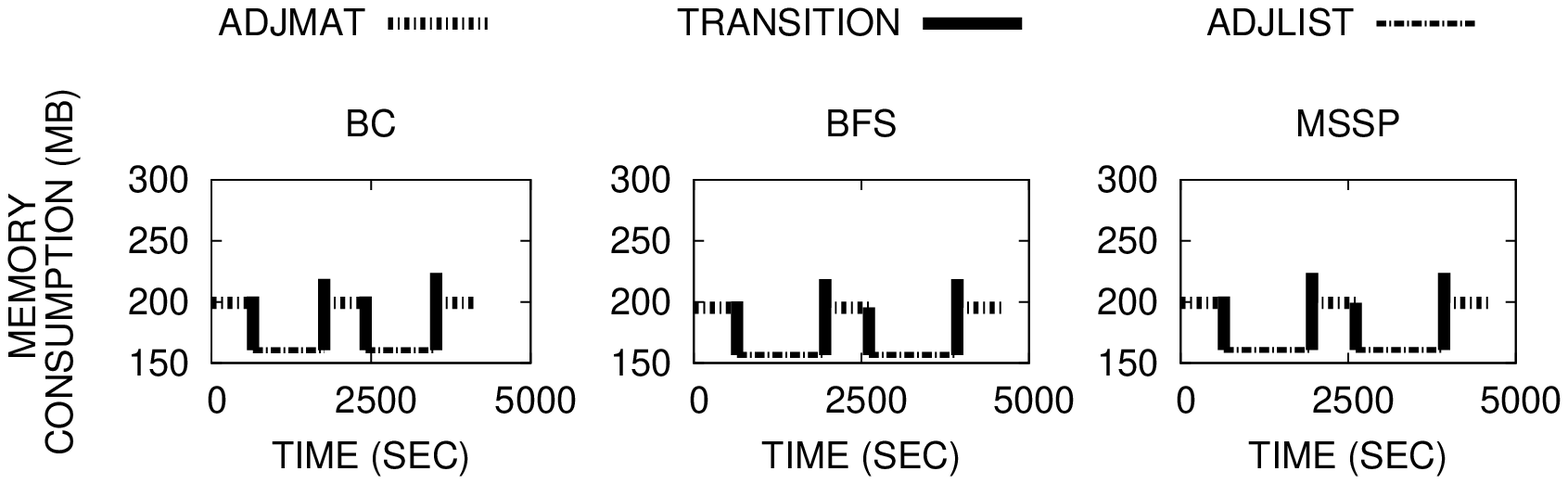}
\vspace{-0.25in}
\hspace{-0.25in}
\includegraphics[width=9cm]{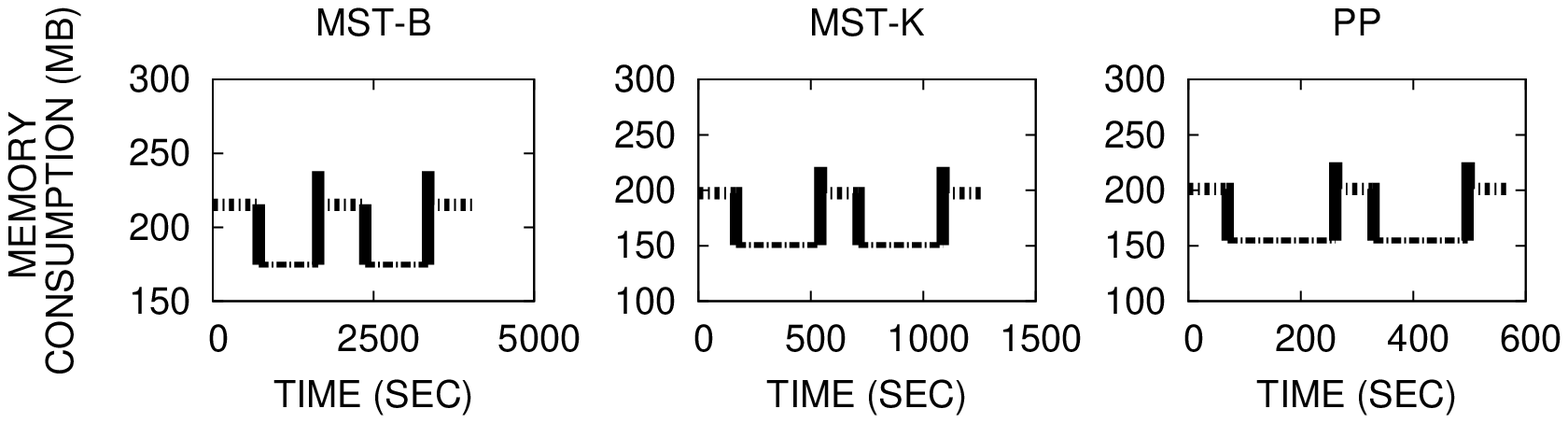}
\caption{System-triggered adaptation for \texttt{wiki-Vote}.}
\vspace{-0.3in}
\label{st-wiki}
\end{figure}

%

\mysubsection{System Triggered Adaptation}
\vspace{0.05in}
In this section we study the second scenario, i.e., when the adaptation is triggered by the system. The graph used for these experiments was {\tt p2p-Gnutella} at 20\%
density. However, we select \texttt{ADJMAT} as the initial data structure representation so that no adaptation was triggered due to the mismatch between the data structure and graph density. Instead we provided the program with a system trigger that forces the program to reduce its memory consumption. This causes 
adaptation to be triggered, and the program to switch from
\texttt{ADJMAT} to \texttt{ADJLIST} representation to save memory. As
expected, the execution takes longer. Since the conversion from one
representation to another can be triggered at any time during a
program's execution, in this study we present data for different
trigger points -- after 25\%, 50\%, and 75\% of total processing. We
controlled the trigger point by tracking the amount of processing that
has been completed. 

The results are presented in
Figure~\ref{st-table}. The execution times of the following versions
are presented: non-adaptive version in the \texttt{ADJMAT}
representation (leftmost bar); three adaptive versions with different
trigger points (middle three bars); and non-adaptive \texttt{ADJLIST}
(rightmost bar). All times are normalized with respect to the time for
non-adaptive \texttt{ADJLIST}. As we can see, the execution time of
the adaptive version is always greater than the non-adaptive
\texttt{ADJMAT} version and less than the non-adaptive
\texttt{ADJLIST} version. In other words, if large amounts of memory
are available for longer duration, the adaptive version yields greater
reduction in execution time over the non-adaptive \texttt{ADJLIST}
version. 

To study the behavior of our approach when there are multiple
transitions, we ran experiments on {\tt wiki-Vote} at 10\% density in the following scenario. For each
benchmark, the execution was started with \texttt{ADJMAT} and then
switched to \texttt{ADJLIST} and vice versa after 20 \%, 40\%, 60\%
and 80\%. We controlled the triggers for memory changes from the
system by tracking the amount of processing that has been
completed. We present the results in Figure~\ref{st-wiki}. We can clearly see that, during a resource crunch
when available memory decreases, our applications adapt to decrease
their memory requirements accordingly, hence running slower; after the
resource crunch is over, our applications re-assume the uncompressed
representation and their performance increases.
\vspace{-0.16in}
\mysubsection{Limitations of Our Approach}
\vspace{0.03in}
First, our approach is only useful when the alternative data structures offer a significant trade-off between memory usage and execution time. For example, for the \emph{agglometric clustering} benchmark, when we tried using two alternate data structures of kd-tree and r-tree, we observed no significant trade-off between memory usage and execution time. Since there is a need to bulk load the data, the kd-tree always outperforms the r-tree. Second, our approach is only useful when the application is sufficiently compute and data intensive to justify the cost of runtime monitoring and transition logic. For example, in the case of the Max Cardinality Bipartite Matching benchmark, although the trade-off exists, the benchmark is not sufficiently compute-intensive to justify the adaptation cost.

\mysection{Related Work}
\vspace{0.01in}
\label{sec_Rel}
There is a large body of work on program transformations applied at compile-time or runtime to enhance 
program performance, which also influences resource usage. Some of these techniques can be used to support adaptation. ContextErlang~\cite{Ghezzi} supports the construction of self-adaptive software using different call back modules. Compiler-enabled adaptation techniques include altering of the contentiousness of an application~\cite{tang,tang2}, which enables co-location of applications without interfering with their performance; data spreading~\cite{sds} migrates the application across multiple cores; adaptive loop transformation~\cite{gupta1} allows a program to execute in more than one way during execution based on runtime information. Multiple applications that are running on multicore systems can significantly impact each other's performance as they must share hardware resources (e.g., last level cache, access paths to memory)~\cite{r5}. The impact of interference on program performance can be predicted and estimated~\cite{r6,r7}, and contention management techniques guided by last level shared cache usage and lock contention have been developed~\cite{r8,r9,r10,r11,r12,r13}.

Huang et al. proposed Self Adaptive Containers~\cite{adaptiveContainer} where they provide the developer with a container library which adjusts the underlying data structure associated with the container to meet Service Level Objectives (SLO); adaptation occurs during SLO violations. Similarly, CoCo~\cite{coco} allows adaptation by switching between Java collections during execution depending on the size of collection. These methods are orthogonal to our approach as they do not have scope for user-defined data structures, and the space-time tradeoff is not taken into consideration.

\mysection{Conclusion}
\label{sec_Conc}
\vspace{0.04in}Graph applications have resource requirements that vary greatly across
runs due to differences in graph characteristics; moreover, the
required memory might not be available due to pressure from co-located
applications. We have observed that data structure choice is crucial for allowing the application to get the best out of available resources. We propose an approach that uses programming and runtime support to allow graph applications to be transformed into adaptive applications by choosing the most appropriate data structure. Experiments with graph-manipulating applications which adapt by switching between data structure representations show that our approach is easy to use on off-the-shelf applications, is effective at performing adaptations, and imposes very little overhead.

\section*{Acknowledgments}
This work was supported in part by NSF grants CCF-0963996 and CCF-1149632.
This research was sponsored by the Army Research Laboratory and was
accomplished under Cooperative Agreement Number W911NF-13-2-0045 (ARL
Cyber Security CRA). The views and conclusions contained in this
document are those of the authors and should not be interpreted as
representing the official policies, either expressed or implied, of
the Army Research Laboratory or the U.S.  Government. The
U.S. Government is authorized to reproduce and distribute reprints for
Government purposes notwithstanding any copyright notation here on.

\balance

\bibliographystyle{abbrv}

\end{document}